# A Comparison Study of the Detection Limit of Omicron SARS-CoV-2 Nucleocapsid by various Rapid Antigen Tests.

*Daniela Dobrynin, Iryna Polischuk, Boaz Pokroy*

*Department of Materials Science and Engineering and the Russell Berrie Nanotechnology Institute*

*Technion – Israel Institute of Technology, Haifa 32000, Israel*

E-mail: bpokroy@technion.ac.il

**Introduction**

Since the first case of COVID-19 disease in Wuhan in December 2019, there is a worldwide struggle to reduce the transmission of acute respiratory syndrome coronavirus 2 (SARS-CoV-2). Many countries worldwide decided to impose local lockdowns in order to reduce person-to-person interactions,[1,2] masks became obligatory especially in closed spaces,[3,4] and there was a general requirement for social distance. However, the most efficient method to reduce continuing spreading of infection among the population, and in the meantime maintain a regular daily life, is early detection of infected contagious people.

Up to now, the most reliable method for SARS-CoV-2 detection is reverse-transcriptase PCR test (RT-PCR). It is possible to detect the virus even if there is only one RNA strand in the sample, and run hundreds of samples simultaneously.[5] This method has a few disadvantages, such as high cost, is time consuming, the need for medical laboratories and skilled staff to perform the test, and the major flaw: the lack of appropriate number of available tests. The latterly prominent Omicron variant (B.1.1.529) and its derivatives have caused a tremendous increase in the number of infected people due to its enhanced transmissibility.[6–8] This all emphasizes the high demand for easy-to-use, cheap and available detection tests.

The solution was found in the development of rapid antigen tests (RATs), which provides a result within 15-30 minutes. The majority of the industrially manufactured RATs are designed on the basis of detection of the nucleocapsid protein, one out of four main structural proteins in the SARS-CoV-2.[9] The antigen test has proved to be a good diagnostic marker since nucleocapsid protein is detectable even after one day from the disease onset.[10] The key role of the nucleocapsid protein is to pack the viral genome into helical complexes named ribonucleocapsid (RNP), which interacts with the membrane protein (M).[11]

Former studies tested the RATs validity by examining PCR-positive COVID-19 patients or their samples with various RATs.[12–16] These studies are very informative, however, there are many factors that might affect the RAT result, such as the diversity in the viral load (VL) or nucleocapsid concentration in each sample, the swab surface, exposure to different variants, date of infection, immune system reaction to the virus, vaccination status, etc. For instance, it was shown that various RATs tend to detect the Delta and Omicron variants at different sensitivites.[15] This means that infected patients who show PCR-positive result might have a negative-RAT result, simply due to the uniqueness of each person and his situation. Moreover, the nasopharyngeal samples contain many other proteins which might affect the detection of the nucleocapsid protein. In previous studies it was shown that there is a significant variation in sensitivity (34.1%-88.1%) between different RATs.[17] Obviously the limit of detection of the

RAT is lower than that of RT-PCR. Nevertheless, the lower the limit of detection of RAT, the earlier one can detect COVID-19 infected patients. This is important in cases when urgent treatment is needed as well as to facilitate early isolation so as not to infect others.

In this study, we compare seven different commercially available RATs for detection of SARS-CoV-2 by a simple direct experiment. In contrast to previous studies in which PCR positive-confirmed patients were tested by RATs with unknown actual virus load in the samples, here we examined the various RATs by testing with various known pre-defined concentrations of Omicron SARS-CoV-2 Nucleocapsid protein solutions. The tests were performed according to each manufacturer's instructions utilizing the RAT solution containing known pre-defined volumes and concentrations of the nucleocapsid protein of the Omicron variant. This method allowed us to determine the detection limit of each RAT while controlling the amount of the added nucleocapsid protein and disregarding all other influencing factors.

**Materials and methods**

Nucleocapsid protein of the Omicron variant was purchased from BPS Bioscience (San Diego, CA, USA). We used seven commercially available antigen tests: (a) Deepblue COVID-19 (SARS-CoV-2) Antigen Test Kit (colloidal gold) (Anhui Deepblue Medical Technology Co. Ltd, Hefei, Anhui, China), (b) Easy Diagnosis COVID-19(SARS-CoV-2) Antigen Test Kit (Wuhan EasyDiagnosis Biomedicine Co., Ltd, Wuhan, Hubei, China), (c) EcoTest COVID-19 TO-GO (Assure Tech. (Hangzhou) Co., Ltd, Hangzhou, China), (d) GenSure COVID-19 Antigen Rapid Test Kit (GenSure Biotech Inc., Shijiazhuang, Hebei, China), (e) Orient Gene Rapid COVID-19 (Antigen) Self-Test (Zhejiang Orient Gene Biotech Co., Ltd, Huzhou, Zhejiang, China), (f) Panbio COVID-19 Antigen Self-Test (Abbott Labratories, Chicago, Illinois, USA), (g) YHLO GLINE-2019-nCoV Ag for self-testing (Shenzhen YHLO Biotech Co., LTD, Shenzhen, China)

The protein was diluted in Phosphate buffer saline (PBS), pH 7.4 (Sigma-Aldrich, St. Louis, MO, USA) to various concentrations: 550, 55, 5.5, 2.75, 1, 0.55 and 0.275 µg mL$^{-1}$. To each test solution, 5 µL from the relevant concentration was added. The tests were performed as described by the various manufacturer instructions. All tested RATs were based on the same concept; the C-line is the control line, and should appear if the RAT was conducted correctly, and the T-line appears if the nucleocapsid protein was detected. Each type of RAT has checked firstly for the highest concentration, and if the result was clearly positive (the T line was bold and clear), we proceeded to test lower concentrations. If the T-line was barely visible, we performed 3 RAT replicates and continued to lower concentration. Once the T-line was not visible at all, we again tested 3 replicates to ensure the negative result, and determine the detection limit. We took a photo of the RATs once the maximum period mentioned in the manufacturer's instructions was reached (Fig S1-8).

**Results and Discussion**

This work provides evaluation of seven commercially available RATs widely used internationally. In order to determine and compare the detection limit of the different RATs, we directly used Omicron nucleocapsid protein solutions in decreasing concentrations. The protein concentration was lowered continuously, until only the control line (C-line) was visible, rather than two lines (C- and T-lines). The same volume of the protein solution (5 µL) was added into the test solution each time, implying that the total amount of the nucleocapsid was identical, yet the final concentration of the protein in the test solution may vary. The performance of the different RATs was further compared eliminating any human factor, such as the manner in which one may conduct the RAT and the actual virus load of the tested patient.

Our results suggest that all RATs were able to detect the nucleocapsid protein using the 5.5 µg mL$^{-1}$ solution, while the lowest detected concentration was 0.55 µg mL$^{-1}$, by (b) and (e) RATs (Fig. 1). It is important to stress, that in both cases, only one out of three RATs showed a very thin T-line. In general, the reduction in the concentration of nucleocapsid protein caused the T-line to be weaker and thinner, as expected. (c), (d) and (g) RATs were able to detect the nucleocapsid until 1 µg mL$^{-1}$ solution, and (f) RAT showed positive results until 2.75 µg mL$^{-1}$ nucleocapsid concentration. These results provide additional evidence to previous studies suggesting that there might be significant variation between different RATs in detecting SARS-CoV-2, and especially the Omicron variant. In this study, we observe up to one order of magnitude of nucleocapsid concentration difference between the several RATs.

As can be seen in Fig. S8, we compared three different RATs while using the same concentration of nucleocapsid protein. The emerged T-lines are not similar, in the case of (f) RAT the T-line is very thick and clear, the T-line observed on (d) RAT is weaker but detectable with the human eye, and in the case of (a) RAT there is no visible T-line. On the other hand, (d) RAT shows better detection capacity than these two RATs.

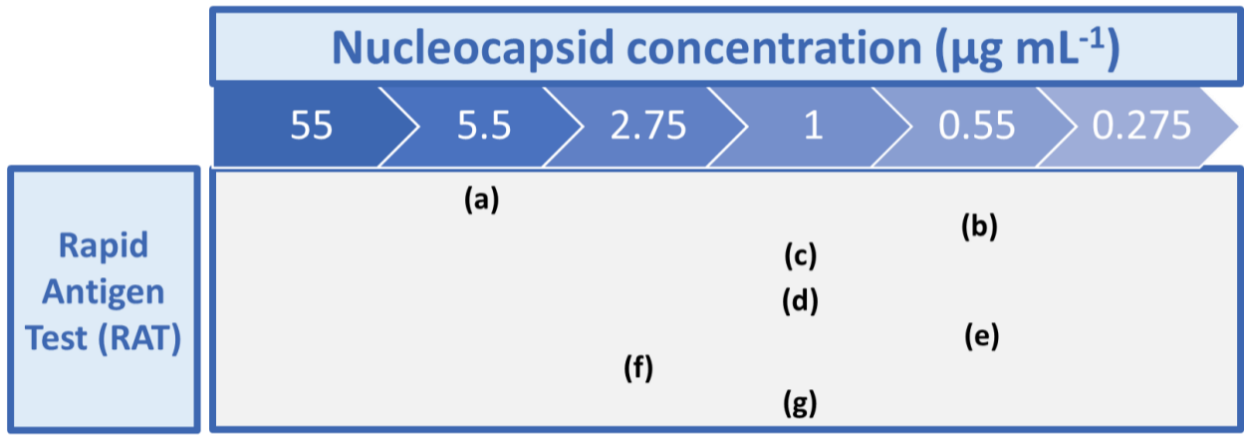

**Figure 1**- The lowest nucleocapsid concentration that was added to the test solution of various RATs, which have shown a positive result (C and T-lines). The RATs are: (a) Deepblue COVID-19 (SARS-CoV-2) Antigen Test Kit (colloidal gold), (b) Easy Diagnosis COVID-19(SARS-CoV-2) Antigen Test Kit (c) EcoTest COVID-19 TO-GO, (d) GenSure COVID-19 Antigen Rapid Test Kit (e) Orient Gene Rapid COVID-19 (Antigen) Self-Test (f) Panbio COVID-19 Antigen Self-Test, (g) YHLO GLINE-2019-nCoV Ag for self-testing.

Previous studies have shown that every twelve nucleocapsid proteins interact with ~800 nucleotides of genomic RNA to produce a RNP complex. Within each virus there are 35-40 RNPs[18,19], meaning that the total amount of nucleocapsid protein varies between 420 to 480 entities according to this model. Assuming that the weight of the protein is 48 KDa, we can translate the amount of protein to the corresponding number of viruses. For instance, 5 µL of 550 µg mL$^{-1}$ nucleocapsid solution is equivalent to 7.19-8.21*10$^{10}$ virions, while the lowest detectable protein concentration of 0.55 µg mL$^{-1}$, corresponds to 7.19-8.21*10$^{7}$ virions. This analysis emphasizes the high amount of virus copies that are needed in order to receive a positive result, in contrast to nucleic acid amplification techniques, which in the presence of only several virions are able to detect an infected person.

The average VL of nasopharyngeal swabs (NS) in infected patients is about ~$10^8$ genome copies per mL, and it varies between $10^5$-$10^{12}$ genome copies per mL.[20,21] Such high variation is possible due to various factors, such as vaccination, variant type, number of days since infection, etc. For instance, the VL of vaccinated patients is lower than non-vaccinated ones, and Delta variant causes higher VL than Omicron variant.[21] This means that higher VL will lead to a higher nucleocapsid amount in the test solution, and a lower probability of having a false-negative result.

Our study has several limitations though. We have consistently added a specific volume from the protein solution to the test solution, while during a real-life testing, the nasal mucus that discharges to the test solution is not identical each time and might be higher. Moreover, here we use pure nucleocapsid protein dissolved in PBS, while the nasal samples contain the whole virus with the nucleocapsid inside it, interacting with the RNA and M protein. Furthermore, there are various additional proteins, such as mucin, and even other microorganisms in the samples. The factors mentioned above might cause changes in the interaction of the nucleocapsid protein with the RAT, in contrast to pure nucleocapsid that we have used, and it might affect the final result. However, the absence of these factors is uniform in all our tested RATs, which allows a good comparison. In addition, as we can see in Fig S1-7, for the lowest detectable protein concentration, sometimes only one out of three replicates showed a positive result. This means that there are also variations between the RATs from the same company. The increase in the number of the tested RATs of each type may provide more reliable detection limits. Yet, the question arises if in a real-life situation, the patient will persist and test himself more than a few times.

In conclusion, our study demonstrates a simple method to evaluate the detection limit of various RATs, using known pre-defined amounts of pure nucleocapsid protein. Here, as previously shown in other studies yet in a new manner, there is an obvious difference in the detection limits of various commercially available RATs, which customers and medical workers need to take into account when performing the RAT.

**Supplementary**

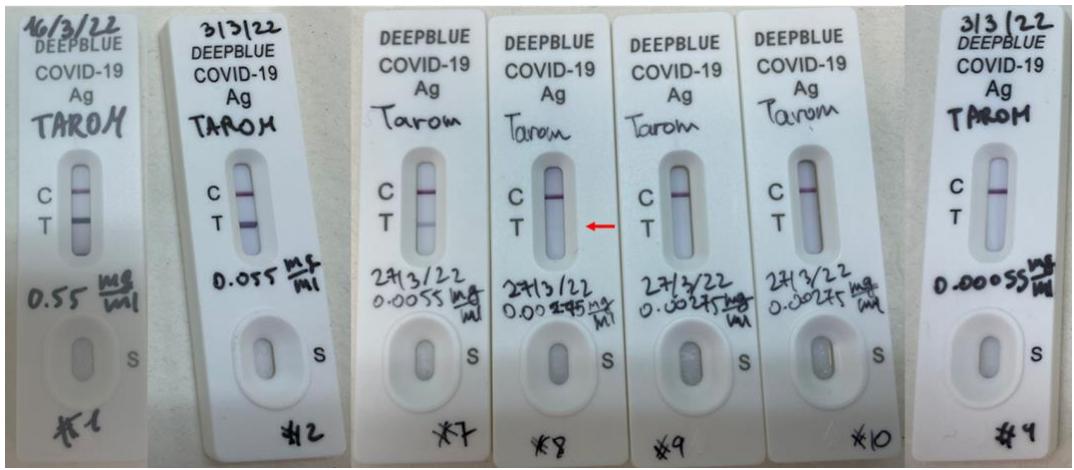

**Figure S1** – "Deepblue COVID-19 (SARS-CoV-2) Antigen Test Kit (colloidal gold)" RATs after reaction with different concentrations of nucleocapsid protein. The lowest concentration that showed positive result is 2.75 µg mL$^{-1}$; only one out of three RAT showed a weak T-line (red arrow).

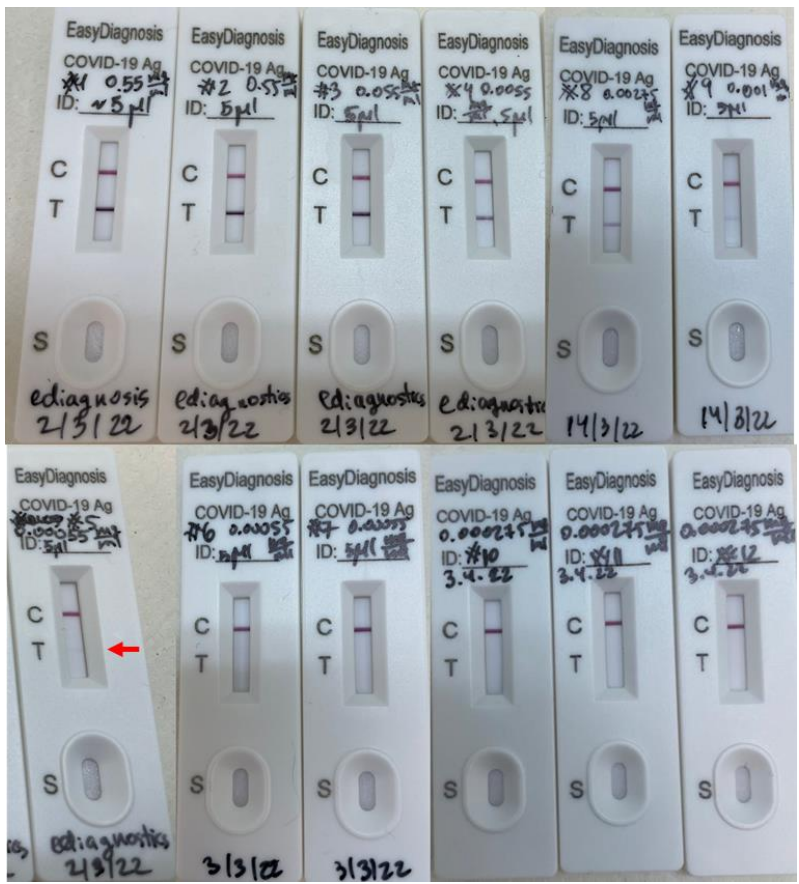

**Figure S2** – "Easy Diagnosis COVID-19(SARS-CoV-2) Antigen Test Kit" RATs after reaction with different concentrations of nucleocapsid protein. The lowest concentration that showed positive result is 0.55 µg mL$^{-1}$; only one out of three RAT showed a weak T-line (red arrow).

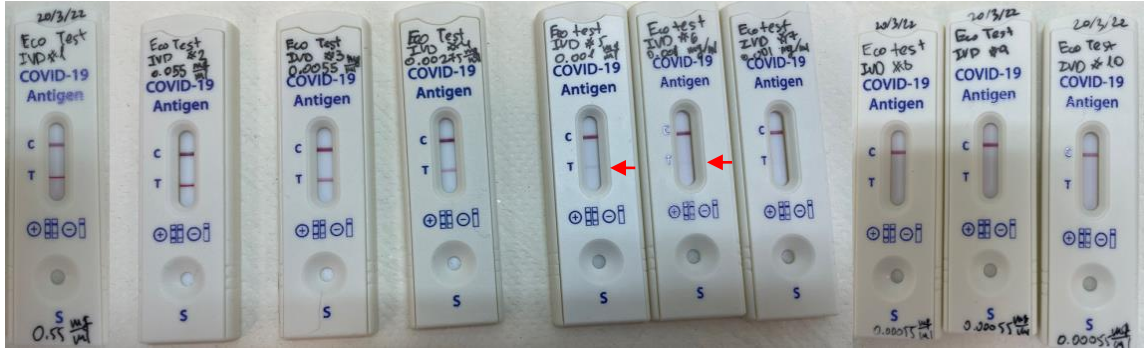

**Figure S3** – "EcoTest COVID-19 TO-GO" RATs after reaction with different concentrations of nucleocapsid protein. The lowest concentration that showed positive result is 1 µg mL$^{-1}$; Two out of three RAT showed a weak T-line (red arrows).

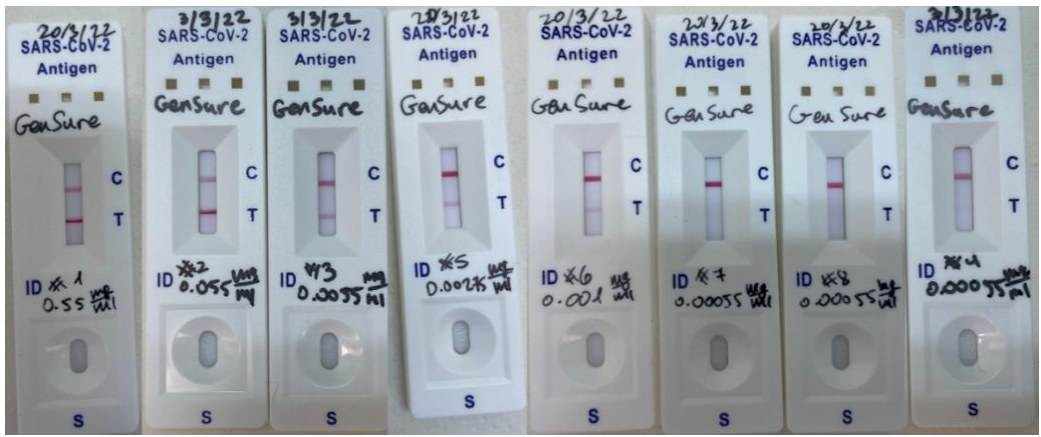

**Figure S4** – "GenSure COVID-19 Antigen Rapid Test Kit" RATs after reaction with different concentrations of nucleocapsid protein. The lowest concentration that showed positive result is 1 µg mL$^{-1}$.

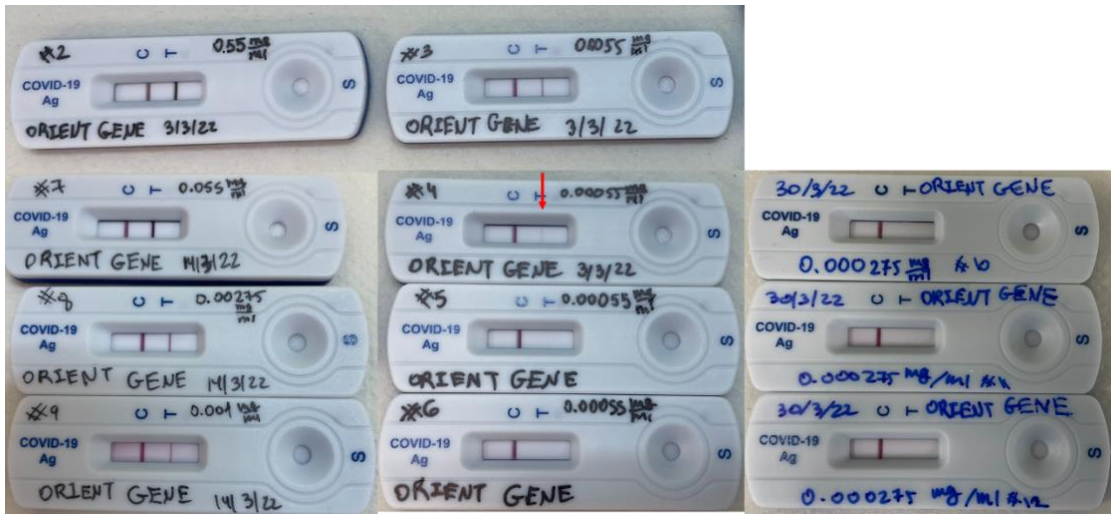

**Figure S5** – "Orient Gene Rapid COVID-19 (Antigen) Self-Test" RATs after reaction with different concentrations of nucleocapsid protein. The lowest concentration that showed positive result is 0.55 µg mL$^{-1}$; Only one out of three RAT showed a T-line (red arrow).

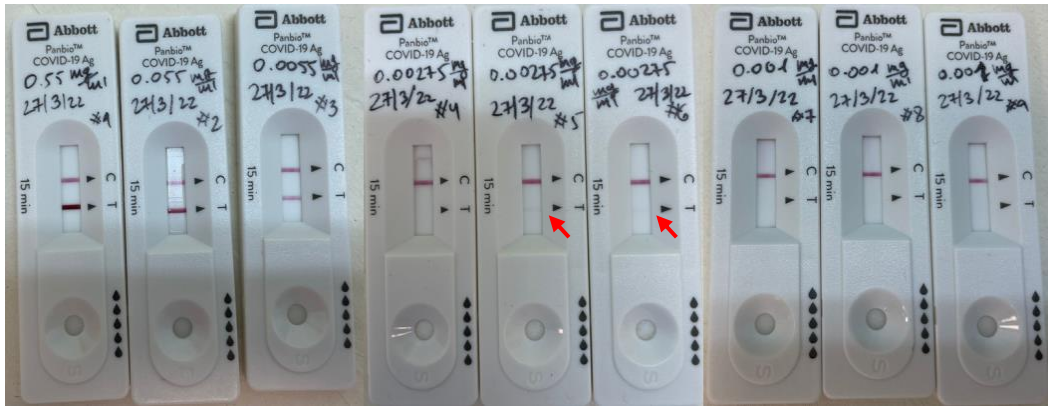

**Figure S6** – "Panbio COVID-19 Antigen Self-Test" RATs after reaction with different concentrations of nucleocapsid protein. The lowest concentration that showed positive result is 2.75 µg mL$^{-1}$; Two out of three RAT showed a T-line (red arrows).

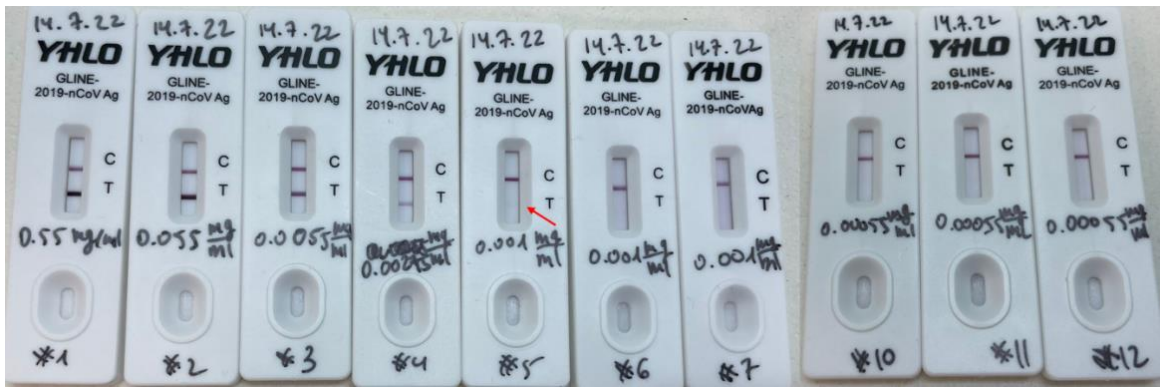

**Figure S7** – "YHLO GLINE-2019-nCoV Ag for self-testing" RATs after reaction with different concentrations of nucleocapsid protein. The lowest concentration that showed positive result is 1 µg mL$^{-1}$; One out of three RAT showed a weak T-line (red arrow).

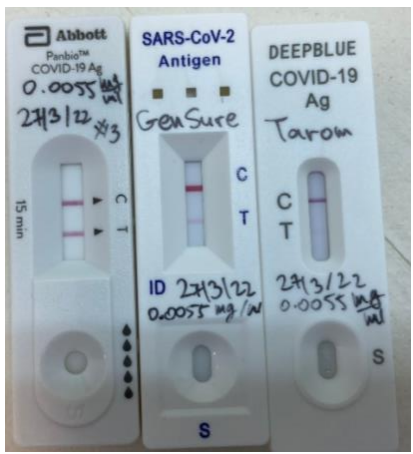

**Figure S8** – Three different RATs after reaction with 5.5 µg mL$^{-1}$ nucleocapsid concentration, at the same day. The T-line is invisible while using "Deepblue COVID-19 (SARS-CoV-2) Antigen Test Kit (colloidal gold)" RAT, intermediate intensity of T-line observed in "GenSure COVID-19 Antigen Rapid Test Kit" RAT, and the "Panbio COVID-19 Antigen Self-Test" RAT shows the thickest T-line.